\documentclass{article} 
\usepackage{iclr2025_conference,times}
\usepackage{amsmath}
\usepackage{algorithm}
\usepackage{algpseudocode}
\usepackage{pifont}
\usepackage{graphicx} 
\usepackage{xcolor}
\usepackage{tcolorbox}
\usepackage{booktabs}
\usepackage{threeparttable}
\usepackage{svg}
\usepackage{array}

\usepackage{amsmath,amsfonts,bm}

\def\eqref#1{equation~\ref{#1}}

\def\1{\bm{1}}

\DeclareMathAlphabet{\mathsfit}{\encodingdefault}{\sfdefault}{m}{sl}
\SetMathAlphabet{\mathsfit}{bold}{\encodingdefault}{\sfdefault}{bx}{n}

\usepackage{hyperref}
\usepackage{url}

\title{o1-Coder: an o1 Replication for Coding}

\iclrfinalcopy

\author{Yuxiang Zhang, Shangxi Wu, Yuqi Yang, Jiangming Shu, Jinlin Xiao, Chao Kong \& Jitao Sang\thanks{Corresponding author.} \\
School of Computer Science and Technology\\
Beijing Jiaotong University\\
Beijing, China\\
\texttt{\{yuxiangzhang, wushangxi, yqyang, jiangmingshu, jinlinx, 23120361, }\\
\texttt{jtsang\}@bjtu.edu.cn} \\
\small
}

\begin{document}

\maketitle

\begin{abstract}
The technical report introduces O1-CODER, an attempt to replicate OpenAI's o1 model with a focus on coding tasks. It integrates reinforcement learning (RL) and Monte Carlo Tree Search (MCTS) to enhance the model's System-2 thinking capabilities. The framework includes training a Test Case Generator (TCG) for standardized code testing, using MCTS to generate code data with reasoning processes, and iteratively fine-tuning the policy model to initially produce pseudocode and then generate the full code. The report also addresses the opportunities and challenges in deploying o1-like models in real-world applications, suggesting transitioning to the System-2 paradigm and highlighting the imperative for world model construction. Updated model progress and experimental results will be reported in subsequent versions. All source code, curated datasets, as well as the derived models are disclosed at \hyperlink{https://github.com/ADaM-BJTU/O1-CODER}{https://github.com/ADaM-BJTU/O1-CODER}.
\end{abstract}

\section{Introduction}
\begin{figure}[t!] 
  \centering 
  \includegraphics[width=0.98\textwidth]{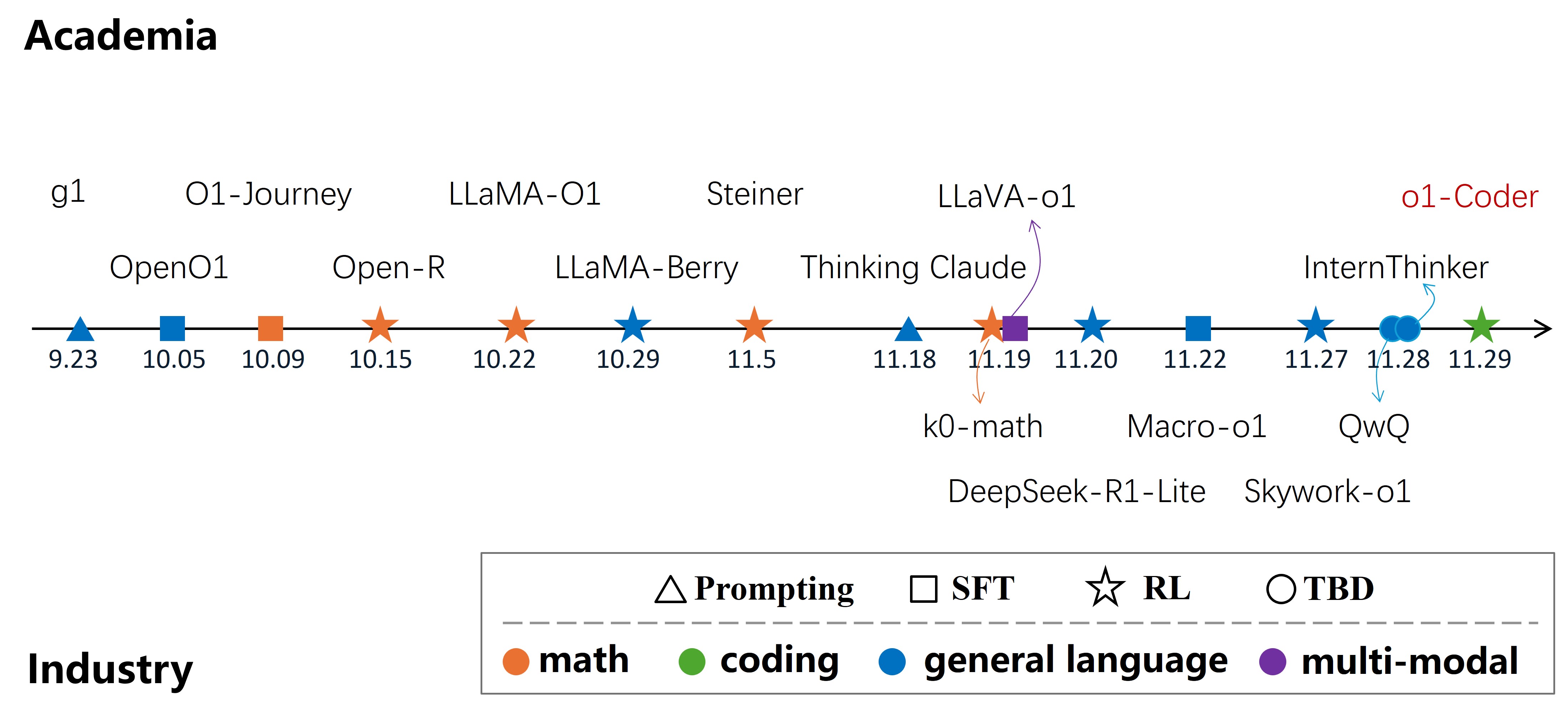} 
  \caption{o1 replication efforts: upper part from academic institutions and open-source communities, and lower part from the industry.} 
  \label{fig:1} 
\end{figure}

OpenAI recently introduced the o1 model~\citep{openai_o1_2024}, which has demonstrated impressive system-2 thinking capabilities. This model represents a significant advancement in AI’s ability to perform complex reasoning tasks that require higher-order cognitive functions. Following its release, numerous analysis and replication efforts have emerged, demonstrating the growing interest in reasoning models. Notable works include g1~\citep{g1_github}, OpenO1~\citep{openo1}, O1-Journey~\citep{o1_journey_2024}, OpenR~\citep{openr_2024}, LLaMA-O1~\citep{llama_o1_2024}, LLaMA-Berry~\citep{llama_berry_2024}, Steiner~\citep{ji2024steiner}, Thinking Claude~\citep{thinkingclaude}, LLaVA-o1~\citep{xu2024llavao1letvisionlanguage}, and several industrial releases such as k0-math, DeepSeek-R1-Lite, Macro-o1~\citep{zhao2024marcoo1openreasoningmodels}, Skywork o1, QwQ~\citep{qwq_blog}, and InternThinker~\citep{internlm_chat} (illustrated in Fig.~\ref{fig:1}). 

Prior to the o1 model, large language models (LLMs) primarily exhibited System-1 capabilities, characterized by fast, intuitive responses. These models were trained on datasets consisting mainly of question-answer $(Q, A)$ pairs, lacking the intermediate reasoning steps that involve deliberate and analytical processing. This stems from the fact that humans rarely record their thought processes on the internet or elsewhere. Traditionally, techniques such as Chain-of-Thought (CoT) prompting were used to guide models in generating step-by-step reasoning before arriving at an answer. A more direct and effective way is to create datasets including the reasoning sequences, e.g., $(Q, ..., S_i, ..., A)$, where $S_i$ represents an individual reasoning step leading to the final answer. Thus, a second approach to enhancing System-2 reasoning capabilities is supervised fine-tuning (SFT). However, most publicly available data is recorded in a question-answer form, and annotating or distilling such data, especially for complex tasks, is both costly and challenging. In this study, we aim to explore in the absence of reasoning process data, and thus opt for the third approach of reinforcement learning (RL).

It is widely believed that o1 addresses the lack of reasoning data by combining reinforcement learning with pretraining. Reinforcement learning is well known for its ability to explore and discover new strategies rather than relying on predefined data in the past decade. Looking back at key developments in machine learning, we can see that deep learning and large-scale pretraining have driven transformations in model architecture and the requirements for labeled data, respectively. In contrast, reinforcement learning addresses a different aspect of transformation on the objective function. In situations where explicit guidance or clear goals are absent, RL exploits exploration to search for new knowledge and solutions. Therefore, combining pretraining with RL creates a powerful synergy of learning and search, where pretraining compresses existing human knowledge, and RL enables the model to explore new possibilities.

We chose coding tasks to explore how to employ RL to generate and refine reasoning data.  
Coding is a typical task that requires System-2 thinking, involving careful, logical, and step-by-step problem-solving.  Moreover, coding can serve as a foundational skill for solving many other complex problems. This report presents our attempt to replicate o1 with a specific focus on coding tasks. The approach integrates RL and Monte Carlo Tree Search (MCTS) to enable self-play, allowing the model to continually generate reasoning data and enhance its System-2 capabilities.

\section{Framework Overview}
There are two main challenges to address for applying self-play RL to code generation. The first challenge is result evaluation, i.e., assessing the quality of the generated code. Unlike tasks such as Go or mathematics, where results can be directly evaluated based on game rules or correct answers, evaluating code requires running the generated code within a testing environment and verifying it against test cases. We cannot assume that code datasets will always provide sufficient test cases. The second challenge involves defining the thinking and search behaviors, i.e., determining the state transition and the granularity of process rewards. For code generation, the key question is how to design the reasoning process and the space of policies to guide the model’s behavior effectively.

To address the first challenge, we propose training a Test Case Generator (TCG), which automatically generates test cases based on the question and the ground-truth code~\footnote{We also propose an alternative approach where test cases are generated based solely on the question. In addition to be capable of utilizing code datasets that only provide questions, it can also be applied during the inference phase, enabling online reasoning without the need for ground-truth code.}. This approach will help build a standardized code testing environment, providing result rewards for reinforcement learning. 

For the second challenge, two possible approaches can be considered. One is ``think before acting", where the model first forms a complete chain-of-thought and then generates the final answer all at once. The other approach, ``think while acting"~\citep{Zelikman2024QuietSTaR}, involves generating parts of the answer while simultaneously reasoning through the task. We chose the former approach in this study. For code generation, this means first thinking through and writing out a detailed pseudocode, which is then used to generate the final executable code. The advantages are two-fold: adaptability, as the same pseudocode can lead to different concrete code implementations; and controllable granularity, as adjusting the level of detail in the pseudocode can control the granularity of the reasoning/search behavior.

\begin{figure}[tbp] 
  \centering 
  \includegraphics[width=0.75\textwidth]{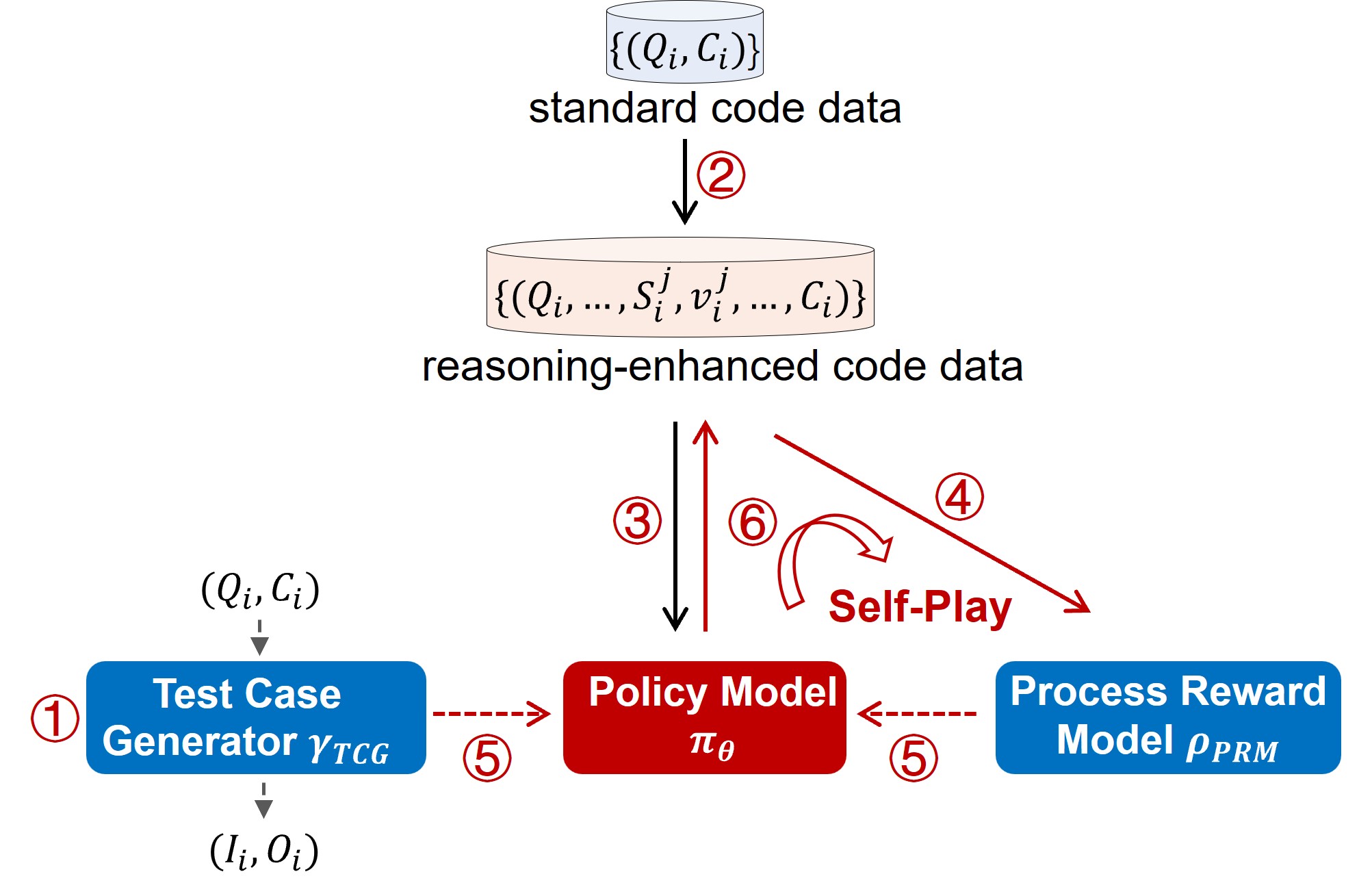} 
  \caption{Self-Play+RL training framework.} 
  \label{fig:2} 
\end{figure}

The outlined framework is provided in Algorithm~\ref{alg:self-play-code-gen}, which consists of six steps. (1) The first step is training the test case generator (TCG) $\gamma_{\text{TCG}}$, which is responsible for automatically generating test cases based on the question. (2) In the second step, we run MCTS on the original code dataset to generate code dataset with reasoning processes $\mathcal{D}_{\text{process}}$, including a validity indicator to distinguish between correct and incorrect reasoning steps. (3) 
Once we have data that includes the reasoning process, the third step is to fine-tune the policy model $\pi_{\theta}$, training it to behave in a ``think before acting" manner. (4) The reasoning process data can also be used to initialize the process reward model (PRM) $\rho_{\text{PRM}}$, which evaluates the quality of reasoning steps. (5) The fifth step is the most crucial: with PRM $\rho_{\text{PRM}}$ providing process rewards and TCG $\gamma_{\text{TCG}}$ providing outcome rewards, the policy model $\pi_{\theta}$ is updated with reinforcement learning. (6) In the 6th step, based on the updated policy model, new reasoning data can be generated. This new data can then be used to fine-tune the PRM again (4th step). Therefore, steps 4, 5, and 6 form an iterative cycle, where self-play continues to drive model improvements. The flow between the six steps is illustrated in Fig.~\ref{fig:2}. The following section will introduce each step in detail.

\begin{algorithm}
\caption{ Self-Play+RL-based Coder Training Framework}
\label{alg:self-play-code-gen}
\begin{algorithmic}[1]

\Require
\Statex $\mathcal{D}_{\text{code}}$: A dataset containing problems $Q_i$ and solution code $C_i$.
\Statex $\pi_{\theta}$: Initial policy model
\Statex $\gamma_{\text{TCG}}$: Test Case Generator(TCG) to create problem-oriented test samples
\Statex $\rho_{\text{PRM}}$: Process Reward Model(PRM) to evaluate the quality of intermediate reasoning steps
\Statex $\phi$: Aggregation function combining result-based and process-based rewards

\Ensure
\Statex Optimized policy model $\pi_{\theta}^*$

\vspace{0.5em}

\Statex \Comment{\textcolor{blue}{\ding{172} Train the Test Case Generator (TCG)}}
\State Train $\gamma_{\text{TCG}}$ on $\mathcal{D}_{\text{code}}$ to maximize diversity and correctness of generated test cases $\{(I_i, O_i)\}$.

\vspace{0.5em}

\Statex \Comment{\textcolor{blue}{\ding{173} Synthesize Reasoning-enhanced Code Dataset}}
\State Based on $\mathcal{D}_{\text{code}}=\{Q_i, C_i\}$, use MCTS to generate $\mathcal{D}_{\text{process}}=\{(Q_i,\cdots, S_i^j,v_i^j,\cdots,C_i')|j=1,\cdots,m\}$, where $S_i^j$ represents a reasoning step and $v_i^j\in \{0, 1\}$ is a validity indicator with $v_i^m=1$ when the generated code pass the test cases.

\vspace{0.5em}

\Statex \Comment{\textcolor{blue}{\ding{174} Finetune the Policy Model}}
\State Finetune $\pi_{\theta}$ with SFT on valid steps  $\mathcal{D}^+_{\text{process}} = \{(Q_i, S_i^j, C_i') \mid (Q_i, S_i^j, v_i^j, C_i') \in \mathcal{D}_{\text{process}}, \mathbb{I}(C_i^{\prime})=1\}$.

\vspace{0.5em}

\While{not converged}

    \Statex \Comment{\textcolor{blue}{\ding{175} Initialize/Finetune the Process Reward Model (PRM)}}
\State Train/Finetune $\text{PRM}$ using SFT on $\mathcal{D}_{\text{process}}$ with point-wise loss, or using DPO with pair-wise loss.

    \vspace{0.5em}

\Statex \Comment{\textcolor{blue}{\ding{176} Improve the Policy Model with Reinforcement Learning}}
\State Initialize $r_i = 0$.
\For{$j = 1, 2, \dots, m$}
    \State Generate reasoning step $S_i^j \sim \pi_{\theta}(S_i^j \mid Q_i, S_i^{1:j-1})$.
    \State Use $\text{PRM}$ to compute process-based reward $r_i^j = \rho_{\text{PRM}}(Q_i,S_i^{1:j})$.
\EndFor
\State Based on $Q_i$ and the complete reasoning sequence $S_i^{1:m}$, generate the final code $C_i'$.
\State Use $\text{TCG}$ to generate test cases $(I_i, O_i)$ for each problem $Q_i$ with the ground-truth code $C_i$.
\State Execute generated code $C_i'$ on inputs $I_i$ to produce outputs $O_i'$.
\State Compute result-based reward:
\[
R_i = \begin{cases} 
\tau_{\text{pass}}, & \text{if } O_i' = O_i, \\
\tau_{\text{fail}}, & \text{otherwise.}
\end{cases}
\]

\State Update $\pi_{\theta}$ using a reinforcement learning method guided by the aggregated reward $\phi(R_i, r_i^{1:m})$.

 \State \Comment{\textcolor{blue}{\ding{177} Generate New Reasoning Data}}
    \State Generate new reasoning data $\mathcal{D}'_{\text{process}}$ using the updated $\pi_{\theta}$.
    \State Update dataset: $\mathcal{D}_{\text{process}} \gets \mathcal{D}_{\text{process}} \cup \mathcal{D}'_{\text{process}}$.
    
\EndWhile

\vspace{0.5em}

\State \Return Optimized policy model $\pi_{\theta}^*$
\end{algorithmic}
\end{algorithm}

\section{Method and Intermediate Results}

\subsection{Test Case Generator Training}
\subsubsection{Objective}

A Test Case Generator is a tool designed to automate the creation of input-output test cases, which plays a critical role in supporting program verification in code generation tasks.

During the training phase, the correctness of the generated code is typically assessed with standard input-output test cases. The pass rate of these test cases serves as a key metric for evaluating the quality of the generated code and acts as an outcome reward signal to guide the training of the policy model. This reward signal helps the model refine its generation strategy, thereby enhancing its capability to produce accurate and functional code.

In the inference phase, when the trained model is tasked with code generation, standard test cases are often not available to verify the correctness of the generated code. The test case generator mitigates this limitation by providing a self-validation mechanism for the policy model, which allows the policy model to evaluate before final generation. As a result, the policy model is able to select the optimal output path based on the validation results.

\subsubsection{Training}
The training process is divided into two distinct phases: Supervised Fine-Tuning (SFT) and Direct Preference Optimization (DPO)~\citep{rafailov2024direct}.
We denote the generator which is not fine-tuned as $\gamma_{\text{TCG}_{base}}$.

The primary objective of the SFT phase is to ensure that the generator’s output adheres to a predefined format, enabling the accurate parsing and extraction of the generated test cases. The training data for this phase is derived from the TACO dataset~\citep{li2023taco}, which follows the format \{$question$, $solution$, $test\_case$\}. 
To standardize the model’s input and output, we developed a template format, as detailed below:
\begin{figure}[H]
\begin{tcolorbox}[colframe=black!75!white, colback=white!95!black, boxrule=0.75mm, arc=5mm, outer arc=5mm, title=Template format for TCG SFT]
\#\#\# Instruction

Please complete the task in the code part and generate some test case in the test part that can be used to test the quality of the generated code.

\#\#\# Problem
\begin{verbatim}
{question}
\end{verbatim}
\#\#\# Code Part
\begin{verbatim}
{randomly select one solution from the provided solutions}
\end{verbatim}
\#\#\# Test Part
\begin{verbatim}
[Generate 3 test cases here to validate the code]
\end{verbatim}
\begin{verbatim}
{sample 3 test_cases with each formatted as input and output}
\end{verbatim}
\end{tcolorbox}
\caption{Template format for TCG SFT}
\label{fig:template for SFT}
\end{figure}
The generator is denoted as $\gamma_{\text{TCG}_{sft}}$ after SFT.

The goal of the DPO phase is to guide the model in generating test cases that align with specific preferences, thereby enhancing both the performance and reliability of the test case generator. In this study, we employ the DPO method with artificially constructed sample pairs to improve the model's ability to align with desired preferences by constructing a preference dataset.
Our DPO fine-tuning relies on a pre-constructed preference dataset \( D_{pref} = \{x, y_{w}, y_{l}\}\), where \( x \) is prompt that includes instruction, question, and code; \( y_{w} \) is positive example, i.e., test cases that align with the preference; and \( y_{l} \) is negative example, i.e., test cases that do not align with the preference.
We adopt the following rules to construct preference data: for \( y_{w} \), we directly use the three sampled test cases that are completely matched as positive examples; for \( y_{l} \), we shuffle the outputs of the three sampled test cases and then concatenate the original inputs so that the input-output pairs of the three test cases do not completely match, and use the three incompletely matched test cases as negative examples.
The training objective aims to optimize \(\gamma_{\text{TCG}_{\theta}} \) based on initial SFT model $\gamma_{\text{TCG}_{sft}}$, while incorporating implicit reward modeling with the reference model \(\gamma_{\text{TCG}_{\mathrm{ref}}}\), which represents the initial SFT model $\gamma_{\text{TCG}_{sft}}$. The objective function is as follows:
\begin{equation}
\resizebox{0.9\hsize}{!}{$
\mathcal{L}_{\text{DPO}}(\gamma_{\text{TCG}_{\theta}};\gamma_{\text{TCG}_{\mathrm{ref}}}) = -\mathbb{E}_{(x, y_w, y_l) \sim D_{pref}} \left[\log\sigma\left(\beta\log\frac{\gamma_{\text{TCG}_{\theta}}(y_{w}\mid x)}{\gamma_{\text{TCG}_{\mathrm{ref}}}(y_{w}\mid x)}-\beta\log\frac{\gamma_{\text{TCG}_{\theta}}(y_{l}\mid x)}{\gamma_{\text{TCG}_{\mathrm{ref}}}(y_{l}\mid x)}\right)\right],
$}
\end{equation}
where \( \sigma(x) \) is the sigmoid function and \( \beta \) represents a scaling factor used to adjust the contrast strength between the positive and negative examples during training.
The generator is denoted as  $\gamma_{\text{TCG}_{dpo}}$ after DPO, which represents the final generator $\gamma_{\text{TCG}}$.

\subsubsection{Experiments}
We utilize DeepSeek-1.3B-Instruct~\citep{guo2024deepseek} as the base model for the test case generator, followed by SFT and DPO. The fine-tuning phase employs QLoRA technology~\citep{NEURIPS2023_1feb8787} with a rank parameter \( r = 1 \) to adapt the following modules: \( q\_{proj}, o\_{proj}, k\_{proj}, v\_{proj}, gate\_{proj}, up\_{proj}, down\_{proj} \). The learning rate is set to \( 5 \times 10^{-4} \) to balance training stability and convergence speed. The training data is derived from a subset of the TACO train dataset, which adheres to the ACM competition format and contains approximately 10,000 samples. Similarly, the test data is obtained from a subset of the TACO test dataset, also conforming to the ICPC competition format, and consists of 314 samples.

We tested the quality of the generated test cases at different stages of the TACO test. After the SFT phase, the pass rate of test cases generated by $\gamma_{\text{TCG}_{sft}}$ on the standard code was 80.8\%, demonstrating the generator's ability to efficiently produce test cases following preliminary fine-tuning. Furthermore, $\gamma_{\text{TCG}_{dpo}}$ achieved a performance of 89.2\%, reflecting an notable improvement compared to $\gamma_{\text{TCG}_{sft}}$. This indicates that preference optimization, by refining the model’s decision-making process, significantly enhanced the generator’s ability to produce more reliable test cases.

In practical scenarios, the generator’s performance has generally met the requirements for assessing code correctness. Looking ahead, we plan to incorporate the test case generator as an outcome verifier during the inference process. This approach aims to ensure the correctness of generated outputs by validating them against dynamically generated test cases,
enabling more robust inference-time search for code generation. 

Additionally, we are considering the incorporation of self-play in the TCG’s training. In this setup, the policy model would generate code intended to pass the test cases produced by the TCG, while the TCG would aim to generate progressively more challenging test cases. This adversarial interaction could foster mutual improvements in both the policy model and the test case generator.

\begin{figure}[tbh]
\begin{tcolorbox}[colframe=black!75!white, colback=white!95!black, boxrule=0.75mm, arc=5mm, outer arc=5mm, title=Pseudocode Prompt]
\#\#\# Instruction

Please refer to the given task description and provide a thought process in the form of step-by-step pseudocode refinement.
\\
\\
A curious user has approached you with a programming question. You should give step-by-step solutions to the user's questions. For each step you can choose one of the following three actions:
\\
\\
\texttt{<}Action 1\texttt{>} Defining algorithm Structures Using pseudocode

Description: Outline the core functions and overall structure of the solution without getting into implementation details. Define inputs, outputs, and the main tasks each function will perform.
\\
\\
\texttt{<}Action 2\texttt{>} Refine part of the pseudocode

Description: Add more details to the pseudocode, specifying the exact steps, logic, and operations each function will carry out. This prepares the pseudocode for actual coding.
\\
\\
\texttt{<}Action 3\texttt{>} Generate python code from the pseudocode

Description: Translate the refined pseudocode into executable Python code, making sure to handle inputs, outputs, and ensure correctness in the implementation.    
\\
\\
Note:

- You can choose one of the three actions for each step.

- Provide a detailed explanation of the reasoning behind each step.

- Try to refer to the reference code as much as possible, but you can also modify it if needed (e.g. change variable names, add some comments, etc.).
\\
\\
\#\#\# Examples
\begin{verbatim}
{examples}
\end{verbatim}
\vspace{1em}
\#\#\# Question
\begin{verbatim}
{question}
\end{verbatim}
\end{tcolorbox}
\caption{Pseudocode Prompt for Step-by-Step Refinement}
\label{fig:pseudocode_prompt}
\end{figure}

\subsection{Reasoning-enhanced Code Data Synthesis} \label{sec:data-synthesis}
\subsubsection{Pseudocode-based Reasoning Process}

%
The definition of the reasoning process is crucial. As mentioned in the \textit{Introduction}, we explore a pseudocode-based MCTS approach designed to guide large language models in deep reasoning for complex code tasks. Pseudocode, serving as an intermediate representation between natural language descriptions and actual code, offers a more abstract and concise way to express the logical flow of algorithms or programs. To integrate pseudocode reasoning into step-level Chain-of-Thought (CoT), as illustrated in Fig.~\ref{fig:pseudocode_prompt}, we define three key behavioral actions infused with pseudocode reasoning:

\begin{itemize}
    \item \textit{Action 1: Defining Algorithm Structures using Pseudocode}: In this action, the model outlines the structure and interface of the main functions, without delving into implementation details. The aim is to enable the model to grasp the overall task structure, including the inputs, outputs, and core functionalities of each primary function.

    \item \textit{Action 2: Refining the Pseudocode}: In this action, the model iteratively refines the pseudocode defined in Action 1, progressively clarifying the steps, logic, and operations of each function in preparation for the final code implementation.

\item \textit{Action 3: Generating Code from the Pseudocode}: The goal of this action is to accurately translate the structure and logic of the pseudocode into executable code, ensuring that the generated code meets the task requirements.
\end{itemize}

\begin{figure}[!tbp]
\centering 
  \includegraphics[width=0.75\textwidth]{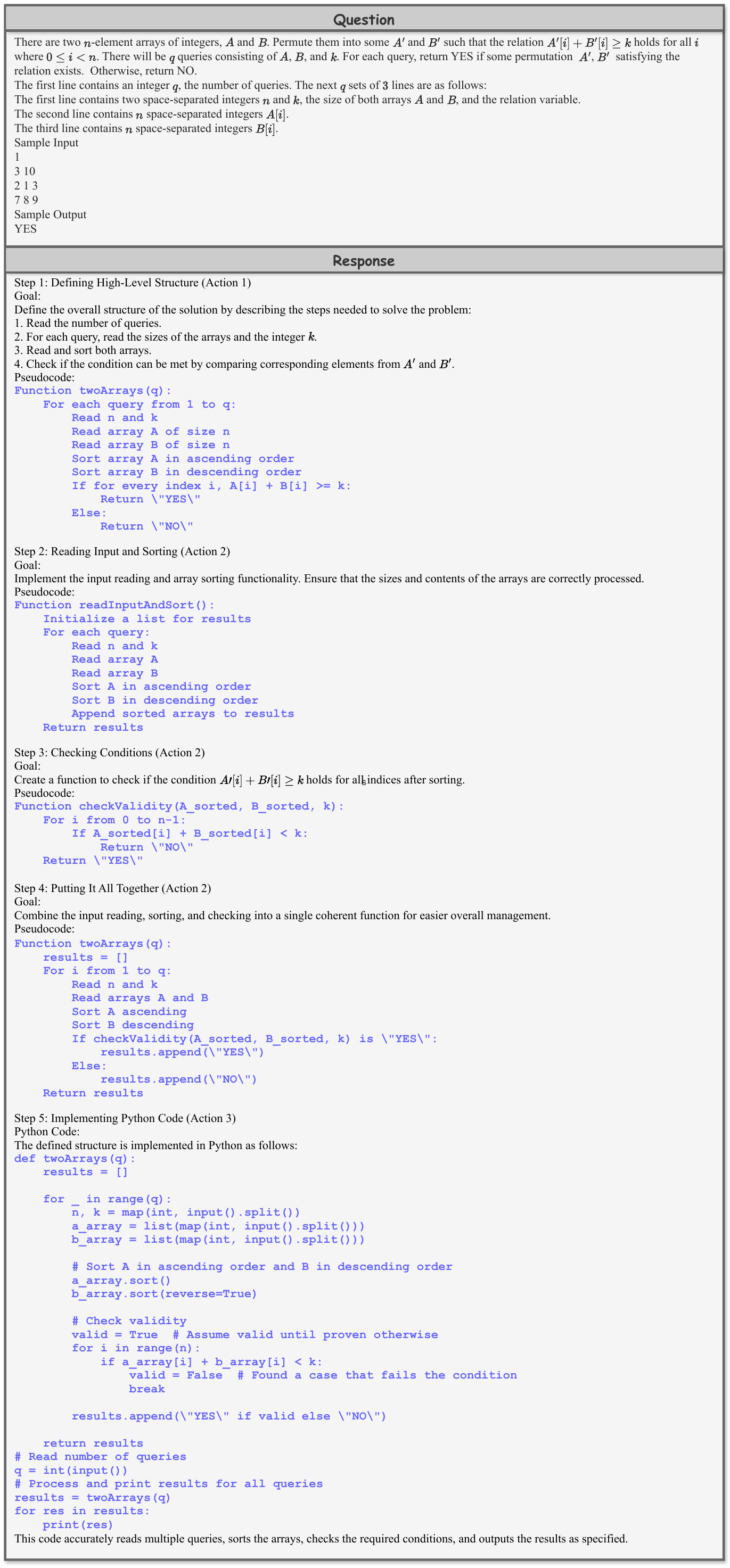}
  \caption{Generated example code with pseudocode CoT} 
  \label{fig:6} 
\end{figure}

These actions ensure that the model employs pseudocode as a cognitive tool during the reasoning process, enhancing its reasoning capability for complex code generation tasks.
It is important to note that these three actions do not imply that the reasoning chain is limited to only these steps. As demonstrated in Fig.~\ref{fig:6}, the model may need to repeatedly invoke Action 2 throughout the reasoning process to iteratively refine the pseudocode until it is sufficiently developed for the final code generation.

To evaluate the effectiveness of the step-level CoT with pseudocode reasoning, we conducted experiments using the Qwen series of open-source models~\citep{yang2024qwen2} and the Mostly Basic Python Problems (MBPP) dataset~\citep{austin2021program} as the benchmark. In the experiment, we employed a sampling strategy based on Monte Carlo Tree Search (MCTS) and compared Pass@1 for regular CoT and CoT with pseudocode reasoning, as well as the Average Sampling Pass Rate (ASPR) of the last step on the correct reasoning path. Our results indicate that incorporating pseudocode significantly improves the quality of the generated code when the reasoning is correct.

Table~\ref{table:1} presents the results. While the Pass@1 metric generally decreases with pseudocode-based reasoning, we observed a significant increase in ASPR, indicating that pseudocode enhances the overall reasoning process, particularly in refining the path toward the correct final output. This suggests that accurate pseudocode highly contributes to the final correct code. However, vanilla LLMs still face challenges in generating effective pseudocode, which is precisely the goal of the subsequent SFT initialization and Self-Play+RL enhancement.

\begin{table}[t]
\centering
    \small
    \resizebox{\textwidth}{!}{
    \begin{tabular}{c c c| c c| c c| c c}
    \toprule[1.5pt]
    Model & \multicolumn{2}{c}{Qwen2.5-1.5B} & \multicolumn{2}{c}{Qwen2.5-3B} & \multicolumn{2}{c}{Qwen2.5-7B} & \multicolumn{2}{c}{Qwen2.5-Coder-7B} \\ 
    \cmidrule(lr){2-3} \cmidrule(lr){4-5} \cmidrule(lr){6-7} \cmidrule(lr){8-9}
     & Vanilla & Pseudocode & Vanilla & Pseudocode & Vanilla & Pseudocode & Vanilla & Pseudocode \\ 
    \midrule
    Pass@1(\%) & \textbf{55.8} & 46.7\textcolor{red}{\textbf{(-9.1)}} & \textbf{56.3} & 51.3\textcolor{red}{\textbf{(-5.0)}} & \textbf{59.8} & 50.1\textcolor{red}{\textbf{(-9.7)}} & 57.7 & \textbf{58.2}\textcolor{green}{\textbf{(+0.5)}} \\ 
    ASPR(\%) & 49.9 & \textbf{54.5}\textcolor{green}{\textbf{(+4.6)}} & 52.0 & \textbf{70.6}\textcolor{green}{\textbf{(+18.6)}} & 66.4 & \textbf{78.1}\textcolor{green}{\textbf{(+11.7)}} & 49.3 & \textbf{74.9}\textcolor{green}{\textbf{(+25.6)}} \\ 
    \bottomrule[1.5pt]
    \end{tabular}
    }
  \caption{Pseudocode-based code generation results on the MBPP Benchmark. \textit{Pass@1} indicates the overall pass rate. \textit{ASPR} (Average Sampling Pass Rate) indicates the average success rate of reaching the correct reasoning path on the last step. } 
  \label{table:1}
\end{table}

\subsubsection{Reasoning Process Data Synthesis}

We use Monte Carlo Tree Search (MCTS)~\citep{kocsis2006bandit, feng2023alphazero, qi2024mutual} to construct step-level process reward data in the form of \(\mathcal{D}_{\text{process}} = \{(Q_i, \cdots, S_i^j, v_i^j, \cdots, C_i^{\prime})\}\), where \(v_i^j\) represents the evaluation of the reasoning path up to step \(S_i^j\), and \(C_i^{\prime}\) is the executable code derived from the final step \(S_i^m\). In this process, we employ the standard MCTS rollout strategy for path exploration. For each problem \(Q_i\), we apply the pseudocode prompt strategy defined earlier to guide the reasoning process. When a terminal node \(S_i^m\) is reached, a complete pseudocode reasoning path \((Q_i, S_i^1, \dots, S_i^m)\) is formed. The reward value \(v_i^m\) for the terminal node \(S_i^m\) is computed based on two key metrics:

\begin{itemize}
    \item \textit{Compilation success rate (compile)}: This metric determines whether the generated code can successfully compile. The value \(\textit{compile}\) is binary, with \(\textit{compile} = 1\) indicating success and \(\textit{compile} = 0\) indicating failure.
    \item \textit{Test case pass rate (pass)}: Given a successful compilation, we further evaluate whether the generated code passes the test cases. The pass rate is calculated as \(\textit{pass} = \frac{\text{Num}_{\text{passed}}}{\text{Num}_{\text{test\_case}}}\), where \(\text{Num}_{\text{passed}}\) is the number of passed test cases and \(\text{Num}_{\text{test\_case}}\) is the total number of test cases used for validation.
\end{itemize}

The reward value for the terminal node \(S_i^m\) is calculated as a weighted sum of these two metrics:

\[
v_i^{m} = \alpha \cdot \text{\textit{compile}} + (1 - \alpha) \cdot \text{\textit{pass}},
\]

where \(\alpha\) is a hyperparameter controlling the relative importance of compilation success and test pass rate.

Once the reward value \(v_i^m\) is computed for the terminal node, we backpropagate this value to all preceding nodes along the path, assigning a reward value \(v_i^j\) to each step \((S_i^j, v_i^j)\). Due to the multiple rollouts in the MCTS process, the cumulative reward for a node \(v_i^j\) during backpropagation may exceed 1. Therefore, we normalize the reward values for each node along the path using the following formula to obtain the final step validity value.

When constructing the reasoning process dataset, for each problem \(Q_i\), if a correct answer is found through the search, we are guaranteed to obtain at least one terminal node \((S_i^m, v_i^m)\) with \(v_i^m = 1\). After completing the search, we select the full reasoning path from the correct terminal node \((Q_i, S_i^1, \dots, S_i^m, v_i^m), v_i^m = 1\) to form the initialization dataset for the policy model. This dataset is denoted as:

\[
\mathcal{D}^+_{\text{process}} = \{(Q_i, S_i^j, C_i^{\prime}) \mid (Q_i, S_i^j, v_i^j, C_i^{\prime}) \in \mathcal{D}_{\text{process}}, \mathbb{I}(C_i^{\prime}) = 1\},
\]

where \(\mathbb{I}(\cdot)\) is an indicator function that returns 1 if the generated code \(C_i^{\prime}\) passes all the test cases.

\subsection{Policy Model Initialization}



After completing the reasoning data synthesis tasks described in Section~\ref{sec:data-synthesis}, we use each complete reasoning solution in the dataset to initialize the policy model \(\pi_{\theta}\). This step aims to help \(\pi_{\theta}\) better understand the task requirements and follow the expected action behavior, providing an optimal starting point for subsequent iterative training.

Given the question \(Q_i\), the specific reasoning step content generated by the policy model \(\pi_{\theta}\) at step \(j\) can be expressed as \(\pi_{\theta}(S_i^j \mid Q_i, S_i^{1:j-1})\), where \(S_i^j = (w_1, w_2, \dots, w_k)\). Here, \(S_i^j\) represents the content of a reasoning step, delimited by specific separators, with \(w\) denoting the tokens generated by \(\pi_{\theta}\) at each decoding step. \(S_i^{1:j-1}\) represents the context formed by the outputs of the previous reasoning steps. 

The policy model \(\pi_{\theta}\) is then initialized using the set of verified, correct reasoning solutions \(\mathcal{D}^+_{\text{process}}\). This initialization is performed by optimizing the following training objective:

\begin{equation}
\mathcal{L}_{\text{SFT}} = -\sum\nolimits_{(Q_i, S_i^j, C_i^{\prime}) \in \mathcal{D}^+_{\text{process}}} \log \pi_{\theta}(S_i^{1:m} \circ C_i^{\prime} \mid Q_i),
\end{equation}

where $\circ$ denotes the concatenation of the reasoning steps $S_i^{1:m}$ and the final code $C_i^{\prime}$. The initialized policy model $\pi_{\theta}^{\text{SFT}}$ will then serve as the foundation for subsequent training stages.

\subsection{PRM Training}

Given a problem \( Q_i \) and a solution prefix corresponding to the current state, the Process Reward Model (PRM), denoted as \( Q \times S \rightarrow \mathbb{R}^+ \), assigns a reward value to the current step \( S_i^j \) to estimate its contribution to the final answer. Based on the tree search approach used during data synthesis in Section~\ref{sec:data-synthesis}, two formats of data organization can be used for training the process reward model, referred to as point-wise and pair-wise, are described in detail below.

\textbf{Point-wise} In this format, data collected from the search tree are organized as \( D = \{(Q_i, S_i^{1:j-1}, S_i^j, v_i^j) \mid i = 1, 2, \ldots, N\} \), where \( N \) is the number of samples, and \( v_i^j \) represents the value label assigned to step \( S_i^j \) during the tree search process. Depending on the processing method, this label can be used to derive either hard or soft estimates. Following the approach in~\citep{wang-etal-2024-math}, the PRM is trained using the objective:

\begin{equation}
\mathcal{L}_{\text{PRM}}^{\textit{point-wise}} = -\mathbb{E}_{(Q_i, S_i^{1:j-1}, S_i^j, v_i^j) \sim D} \Big[ v_i^j \log \, r(Q_i, S_i^{1:j}) + (1 - v_i^j) \log \Big( 1 - r(Q_i, S_i^{1:j}) \Big) \Big],
\end{equation}
where \( r(Q_i, S_i^{1:j}) \) is the normalized prediction score assigned by the PRM.

\textbf{Pair-wise} In the pair-wise format, for a node \( n^d \) at depth \( d \) of the search tree, with its child nodes represented as \( \sum_i n_i^{d+1} \), preference pair data are organized as \( D_{\text{pair}} = \{(Q_i, S_i^{1:j-1}, S_i^{j_{\text{win}}}, S_i^{j_{\text{lose}}}) \mid i = 1, 2, \ldots, N\} \). Here, \( S_i^{j_{\text{win}}} \) represents the reasoning step that achieved a higher value estimate during the tree search compared to \( S_i^{j_{\text{lose}}} \).

Following the Bradley-Terry model~\citep{bradley1952rank}, the PRM is trained using the following objective:
\begin{equation}
\resizebox{0.9\hsize}{!}{$
\mathcal{L}_{PRM}^{\textit{pair-wise}} = -\mathbb{E}_{(Q_i, S_i^{1:j-1}, S_i^{j_{win}}, S_i^{j_{lose}}) \sim D_{pair}} \Big[ \log \Big( \sigma \big( r(Q_i, S_i^{1:j-1}, S_i^{j_{win}}) - r(Q_i, S_i^{1:j-1}, S_i^{j_{lose}}) \big) \Big) \Big],
$}
\end{equation}

where \( \sigma(x) \) denotes the sigmoid function. Unlike the point-wise setting, the scores \( r \) here are not normalized. This enables the model to focus on learning relative preferences between actions rather than absolute value predictions.

\subsection{RL-based Policy Model Improvement}
We model the code generation task as a language-augmented Markov Decision Process (MDP), formally represented as \(\mathcal{M} = (\mathcal{V}, \mathcal{S}, \mathcal{A}, \mathcal{T}, \mathcal{R}, \phi)\)~\citep{openr_2024, carta2023grounding}.
In this framework, \(\mathcal{V}\) denotes the vocabulary, and \(w \in \mathcal{V}\) represents an individual token generated by the model.
The action space \(\mathcal{A} \subseteq \mathcal{V}^N\) and the state space \(\mathcal{S} \subseteq \mathcal{V}^N\) are sets of token sequences, meaning that both actions and states are sequences of tokens. 
In this framework, $s_0$ represents the question, and the action \(a_i\) is considered a reasoning step (referring to the $S_i$ in algorithm \ref{alg:self-play-code-gen}), which consists of both the type of action and its corresponding chain of thought. 
The state transition function \(\mathcal{T}: \mathcal{S} \times \mathcal{A} \to \mathcal{S}\) defines how the current state \(s_t \in \mathcal{S}\) changes when an action \(a_t \in \mathcal{A}\) is taken. Specifically, the action \(a_t\) appends tokens to the current state, forming a new state \(s_{t+1} = \mathcal{T}(s_t, a_t)\). This process continues until the model generates the final solution. The reward function \(\mathcal{R}: \mathcal{S} \times \mathcal{A} \to \mathbb{R}^+\) evaluates the quality of intermediate steps, such as the reasoning process or generated code fragments. The function \(\phi\) combines process-based and outcome-based rewards to produce a final reward signal.

At each step, the model selects an action \(a_t \in \mathcal{A}\), which transitions the system to a new state \(s_{t+1} = \mathcal{T}(s_t, a_t)\). After executing the action, the model receives a process reward \(r^t = \rho_{\text{PRM}}(s_{t-1}, a_t)\) from PRM. This process repeats until the model either generates the final code or reaches the predefined maximum depth.

Once the model generates the final code or completes the search process, the outcome reward \(R_i\) is evaluated by testing the generated code against a series of test cases. We propose a reward aggregation function that incorporates both time-dependent weights and a discount factor:
\[
\phi(R_i, r_i^{1:m}) = \alpha(t) \cdot R_i + (1 - \alpha(t)) \cdot \frac{1}{m} \sum_{j=1}^{m} \gamma^j r_i^j,
\]
where \(\alpha(t)\) is a time-varying factor that adjusts the balance between the final reward \(R_i\) and the cumulative intermediate rewards \(r_i^{1:m}\) over time. For instance, \(\alpha(t)\) may decrease over time, gradually placing more weight on the intermediate rewards as the model refines its solution, while reducing the emphasis on the final reward as the model approaches the optimal policy. \(r_i^{1:m}\), with \(\alpha(t)\) typically following schedules such as linear or logarithmic decay. The parameter \(\gamma \in [0, 1]\) is the discount factor, which determines the importance of future rewards relative to immediate rewards. The aggregated reward signal is employed to refine the model's policy, typically through the implementation of reinforcement learning algorithms such as PPO~\citep{ziegler2019fine} and iterative DPO\citep{rafailov2024direct}.

With this setup, we define a reinforcement learning environment tailored for the code generation task. The model's actions are driven by both process-based rewards, which encourage intermediate reasoning steps, and outcome-based rewards, which reflect the correctness of the final code. This dual reward structure helps the model improve its code generation ability over time.

\subsection{New Reasoning Data Generation and Self-Play}

In step 6, the updated policy model \( \pi_{\theta} \) is used to generate new reasoning data, denoted as \( \mathcal{D}'_{\text{process}} \). This data is created by reasoning through new problem instances \( Q_i \), generating step-by-step reasoning paths \( \{S_i^1, S_i^2, \dots, S_i^m\} \), with each path culminating in a final code output \( C_i^{\prime} \). The reasoning steps are generated iteratively, where each step \( S_i^j \) is conditioned on the previous steps.

Once the new reasoning data is generated, it is added to the existing dataset \( \mathcal{D}_{\text{process}} \) to form an updated dataset \( \mathcal{D}_{\text{process}} \gets \mathcal{D}_{\text{process}} \cup \mathcal{D}'_{\text{process}} \). This update increases the diversity and quality of the reasoning examples, providing more comprehensive training material for subsequent steps. 

This new data generation process enables the iterative self-play training loop. After adding the new reasoning data, the model undergoes further fine-tuning, starting with updating PRM as described in the 4th step. The PRM, in turn, adjusts the policy model with RL described in the 5th step. This iterative cycle of data generation, reward model updating, and policy improvement ensures sustained improvement in the system’s reasoning ability.


\section{Discussions}
\subsection{Bitter Lesson: Data is All You Need}
\begin{figure}[tbp] 
  \centering 
  \includegraphics[width=0.95\textwidth]{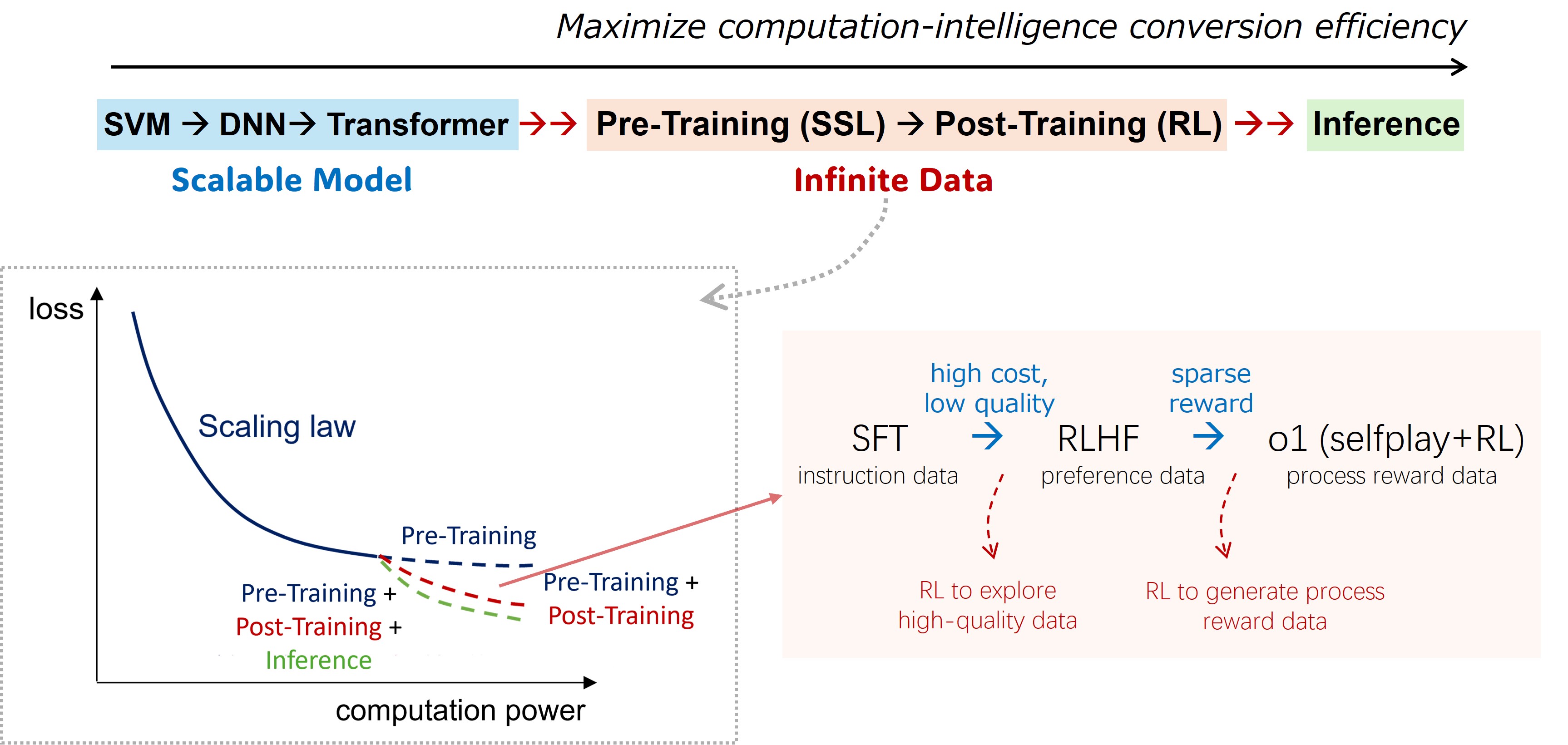} 
  \caption{The trend towards maximizing computation-intelligence conversion efficiency.} 
  \label{fig:3} 
\end{figure}

Over the last decade, the AI field has been developing along a central line towards maximizing computation-intelligence conversion efficiency~\citep{chung2024dont}, which is to efficiently convert the ever-increasing computing power into higher intelligence levels. Along this line, as illustrated at the top of Fig.~\ref{fig:3}, early advancements prioritized improvements on the model side: from SVM to DNN and then to Transformer, scalable model architectures were designed to fully leverage computational power. 

In recent years, the focus has shifted towards the data side. Techniques such as Semi-Supervised Learning (SSL) in pre-training and Reinforcement Learning (RL) in post-training have aimed to harness natural and synthesized data more effectively. The o1 model continues this line. It moves from SFT, which leverages high-quality supervised data, to RLHF, which utilizes environmental feedback to access theoretically unlimited data, and finally to o1’s innovative approach of supervising the generation process through reward signals derived from the generated reasoning process itself.

This progression suggests that, with Transformer architectures now capable of scaling to handle vast amounts of data and training models of sufficient size, the only remaining challenge converges to acquiring adequate data. One approach is to collect data wherever it is lacking, such as reasoning data for system-2 abilities or physical world trajectories for embodied intelligence. Another approach is to explore data types that do not yet exist in the human world, which requires further exploration of techniques like RL and Self-Play.

\subsection{Sweet Lesson: Beyond Human Data}
A common criticism of LLM is its reliance on existing human-recorded data, which inherently limits its potential. As Wittgenstein stated, ``The limits of my language mean the limits of my world.'' The finite scope and depth of human language records constrain the cognitive capabilities of LLMs. However, the success of o1 demonstrates that we can now explore the underlying thought processes behind these recorded data through RL. This advancement signifies a pivotal shift in AI development, moving from mere imitation of human language to the autonomous generation of novel cognitive processes.

More interestingly, these thought process data do not necessarily be confined to natural language. As highlighted in a recent Nature paper, ``language serves primarily as a tool for communication rather than the essence of thought.''~\citep{Fedorenko2024} In our observations, some of the thought chains generated by o1 contain nonsensical text, suggesting that the thinking tokens may not correspond to discrete natural language words. If the model has developed itself a more efficient form of internal representation for thinking, this will significantly elevate the efficiency of thought processes and problem-solving mechanisms, not only transcending the limitations imposed by human language data but also further unlocking the potential of model capabilities.

\subsection{Opportunities: System1 + X to System2 + X}
The self-play RL framework provides a viable solution for exploring underlying data, which opens up the possibility of exploring System-2 solutions for many tasks that were previously reliant on System 1 capabilities. By integrating more thoughtful, step-by-step processes into task execution, we believe that this approach can yield positive results across a wide range of domains~\citep{Kant2024EquitableAccess, ganapini2021thinkingfastslowai, valmeekam2024planningstrawberryfieldsevaluating, lowe20242reasoningcapabilitiesnigh}. Tasks traditionally solved using System 1 capabilities, such as reward modeling~\citep{mahan2024generativerewardmodels}, machine translation~\citep{zhao2024marcoo1openreasoningmodels}, retrieval-augmented generation (RAG)~\citep{li2024alr2}, and multimodal QA~\citep{Islam2024Are}, have already benefited from the deeper reasoning capabilities enabled by System-2 thinking.

The o1 model's system card demonstrates notable improvements in model safety. Inspired by this, we have recently explored the concept of \textit{System-2 Alignment}, which involves guiding models to thoroughly evaluate inputs, consider potential risks, and correct biases in their reasoning~\citep{wang2024dontcommandcultivateexploratory}. We introduced three methods to realize System-2 alignment: prompting, supervised fine-tuning, and reinforcement learning with process supervision. We are applying the Self-Play+RL framework presented in this report to System-2 alignment, aiming to further enhance the model's ability to think deliberately and reduce vulnerabilities in complex scenarios.

\subsection{Challenges: World Model Encoding}
The released o1-preview and o1-mini currently lack multimodal capabilities and functional call features, which are claimed by OpenAI to be included in its complete version. Beyond multimodal and functional call, another critical feature for improvement in o1-like inference models is the optimization of inference time. This includes enhancing inference efficiency—achieving higher performance per unit of time—and enabling adaptive inference time adjustments. Specifically, this involves dynamically adjusting the System 2 reasoning process based on task complexity and achieving a more human-like ability to seamlessly switch between System 1 and System 2 reasoning modes.

For o1-like inference models to be deployed across broader real-world applications, two major challenges need to be addressed, both involving with the RL environments. The first challenge concerns reward function generalization. This has been already discussed in the community. For example, leveraging the enhanced ability of inference models to understand high-level natural instructions, approaches like Constitutional AI~\citep{bai2022constitutional} might directly define reward functions in natural language. Alternative strategy focuses on improving coding capability and transforming the other tasks into coding problems for resolution. 

Another less mentioned challenge concerns environment state update during planning. Unlike classic model-free RL methods without planning, such as Q-learning, where state transitions are not explicitly modeled, o1-like planning models rely on behavior simulation and forward search, requiring knowledge of the updated state following an action. This shifts the paradigm towards model-based RL. Fortunately, in well-defined tasks such as programming, mathematics, and Go, the environment dynamics are often deterministic.  For example, the world models of Go and other board games can be described explicitly through rules. For programming and mathematics, large language models inherently embed their world models regarding programming syntax and axiomatic logic. These deterministic environment dynamics allow precise computation of state transition probabilities following specific actions.

However, in many real-world applications, such as device use~\citep{wang2024mobile,wang2024mobile2} and embodied agents, obtaining state updates requires interaction with external environments or simulators. This introduces significant computational and time costs. For example, in device use, behaviors like clicking, inputting, or scrolling must be simulated in a way that involves page rendering, state updates, and sometimes complex backend interactions like network requests. Moreover, o1-like models face the limitation of not being able to perform online behavior simulation during inference, which prevents the model from validating or correcting its actions by returning to a previous state. This leads to inability to backtrack and refine decisions.

Therefore, one of the key directions is to attempt explicit modeling of the environment by developing a world model for  state transition prediction. The world model takes as input the current and past states as well as actions, and produces the next state as output. This allows the agent to interact with its internal world model, rather than directly with the real environment or a simulator. However, since accurately building such world models is very difficult, world models have typically been applied to environments where the dynamics are relatively simple and well-understood. The good news is, the recent rapid advancements in interactive content generation~\citep{parkerholder2024genie2} and generative games~\citep{generativegame} offer promising progress that could facilitate more accurate and practical environment modeling for planning-based reasoning in real-world applications.

\vspace{3mm}
\textbf{Prospects.} \hspace{1mm} The o1 model is clearly influenced by AlphaGo: AlphaGo utilized imitation learning to initialize the policy network, reinforcement learning to fine-tune the policy and learn the value network, and MCTS as an online search strategy, which parallels LLM's pre-training, post-training, and inference. AlphaGoZero took a more advanced approach by not relying on historical data, which exactly mirrors current trends in LLM development increasingly emphasizing the post-training stage. If we follow their subsequent evolution, we can anticipate similar developments in o1-like reasoning models. 

After AlphaGoZero, the Alpha series first developed towards generalization: AlphaZero was applied to Go, Chess, and Shogi. To further tackle more complex scenarios in Atari video games, MuZero requires a dedicated model to handle state transitions. Its approach involves simultaneously updating a world model and a reward model, enabling model-based planning in a latent space rather than relying on explicit environmental observations. In analogy, selecting a compact state representation and constructing an effective world model to support efficient planning are key to applying o1-like models in real-world scenarios for solving long-horizon reasoning tasks. Interestingly, just a week after the release of o1, 1X, the robotics company backed by OpenAI, unveiled its world model project. This initiative aims to develop a predictive framework to simulate and anticipate the outcomes of actions in real-world environments. It highly envisions the potential applications of o1-like reasoning models in advancing embodied intelligence.

\section*{Acknowledgements}
We thank Yuhang Wang and Jing Zhang for their fruitful discussions and participation.

\bibliography{iclr2025_conference}
\bibliographystyle{iclr2025_conference}

\end{document}